# A parity-breaking electronic nematic phase transition in the spin-orbit coupled metal $Cd_2Re_2O_7$

**Authors:** J. W. Harter[1,2], Z. Y. Zhao[3,4], J.-Q. Yan[3,5], D. G. Mandrus[3,5], D. Hsieh[1,2]*

**Affiliations:**

[1]Department of Physics, California Institute of Technology, Pasadena, CA 91125, USA.

[2]Institute for Quantum Information and Matter, California Institute of Technology, Pasadena, CA 91125, USA.

[3]Materials Science and Technology Division, Oak Ridge National Laboratory, Oak Ridge, TN 37831, USA.

[4]Department of Physics and Astronomy, University of Tennessee, Knoxville, TN 37996, USA.

[5]Department of Materials Science and Engineering, University of Tennessee, Knoxville, TN 37996, USA.

*To whom correspondence should be addressed. Email: dhsieh@caltech.edu

**Abstract**:

Strong electron interactions can drive metallic systems toward a variety of well-known symmetry-broken phases, but the instabilities of correlated metals with strong spin-orbit coupling have only recently begun to be explored. We uncovered a multipolar nematic phase of matter in the metallic pyrochlore $Cd_2Re_2O_7$ using spatially resolved second-harmonic optical anisotropy measurements. Like previously discovered electronic nematic phases, this multipolar phase spontaneously breaks rotational symmetry while preserving translational invariance. However, it has the distinguishing property of being odd under spatial inversion, which is allowed only in the presence of spin-orbit coupling. By examining the critical behavior of the multipolar nematic order parameter, we show that it drives the thermal phase transition near 200 kelvin in $Cd_2Re_2O_7$ and induces a parity-breaking lattice distortion as a secondary order.

**Main Text:**

In the presence of strong Coulomb interactions, the fluid of mobile electrons in a metal can spontaneously break the point group symmetries of the underlying crystal lattice, realizing the quantum analogue of a nematic liquid crystal (*1*). Like their classical counterparts, quantum nematic phases generally preserve spatial inversion symmetry and are therefore anisotropic but centrosymmetric fluids. Experimental evidence of such nematic order was first detected in a two-dimensional (2D) GaAs/AlGaAs quantum well interface on the basis of a pronounced resistivity anisotropy between the two principal directions of the underlying square lattice (*2, 3*). Subsequently, similar behavior has been reported in a number of quasi-2D square lattice compounds, including $Sr_3Ru_2O_7$ (*4*), $URu_2Si_2$ (*5*), and several families of both copper- (*6, 7*) and iron-based (*8–11*) high-temperature superconductors, suggesting possible connections between even-parity nematic fluctuations and unconventional *s*- and *d*-wave Cooper pairing (*12*).

Extending earlier work on Fermi liquid instabilities in the *p*-wave spin interaction channel (*13*), it has recently been predicted that correlated metals with strong spin-orbit coupling may realize a fundamentally new class of electronic nematic phases with spontaneously broken spatial inversion symmetry (*14*), including a quantum analogue of the unusual $N_T$ nematic phase discussed in the context of classical bent-core liquid crystals (*15*). Theoretical models have shown that parity-breaking nematic fluctuations can induce odd-parity *p*- or *f*-wave Cooper pairing and thus provide a route to topological superconductivity (*16, 17*). In addition, because inversion symmetry breaking necessarily lifts the spin degeneracy of bulk energy bands in a spin-orbit coupled system, odd-parity nematic order offers a potential mechanism for generating topologically protected Weyl and nodal-line semimetals and for designing highly tunable charge-to-spin current conversion technologies for spintronics applications.

The order parameter for this predicted new class of spin-orbit-coupled parity-breaking electronic nematic phases—so-called "multipolar" nematics (*14*)—can be represented by a symmetric traceless second-rank pseudotensor $Q_{ij}$ that is odd under spatial inversion. This order parameter induces a deformation and spin splitting of the Fermi surface via the spin-orbit interaction Hamiltonian $H_{SO} = \sum_{ij} Q_{ij}\sigma_i k_j$, where $\sigma_i$ are the Pauli matrices and $k_j$ is the crystal momentum (*14, 18*). In a cubic material, for example, this order parameter can have either $E_u$ or $T_{2u}$ symmetry. An example of a spin-polarized Fermi surface distortion induced by $T_{2u}$ multipolar nematic order is shown in Fig. 1A.

The correlated metallic pyrochlore $Cd_2Re_2O_7$ has been proposed as a candidate for hosting multipolar nematic order because of the strong spin-orbit coupling of Re 5*d* valence electrons. Detailed Raman scattering (*19*), x-ray (*20, 21*) and neutron (*22*) diffraction, and optical second-harmonic generation (SHG) (*23*) studies have shown that at critical temperature ($T_c$) ~ 200 K, the material undergoes a continuous phase transition from a centrosymmetric cubic structure (space group $Fd\bar{3}m$) to a noncentrosymmetric tetragonal structure (space group $I\bar{4}m2$) that breaks threefold rotational symmetry about the ⟨111⟩ axis (Fig. 1, B and C). This phase transition has traditionally been attributed to the freezing of a soft phonon mode with $E_u$ symmetry, dominated by the displacement of O(1) atoms (*19, 24, 25*). However, the observation of extremely small changes in lattice parameters (*21, 22*), an anomalous temperature dependence of superlattice Bragg peaks (*20*), and a dramatic reduction in the electronic density of states across $T_c$ (*26–29*) calls this interpretation into question and raises the possibility that a hitherto undetected electronic order is driving the transition.

Unlike the previously studied even-parity nematic phases, multipolar nematic order cannot be experimentally identified by using charge transport anisotropy measurements because the loss of inversion symmetry is manifested in the spin texture of the Fermi surface. Moreover, observations of conventional nematic order have often relied on an alignment of nematic directors by using applied magnetic (*4*) or uniaxial strain (*9*) fields in order to measure macroscopic symmetry-breaking responses, but neither magnetic nor strain fields couple linearly to $Q_{ij}$ because they are parity-even. Nonlinear optical anisotropy measurements can overcome these challenges because they probe the structure of higher-rank susceptibility tensors that contain full point group information (*30*) and can be performed in a spatially resolved manner. Optical SHG is particularly well-suited to identifying odd-parity phases because the leading-order electric dipole contribution, which is described by a third-rank susceptibility tensor relating the incident electric field to the nonlinear polarization induced at twice the incident frequency via the equation $P_i(2\omega) = \chi_{ijk} E_j(\omega) E_k(\omega)$, is only allowed if inversion symmetry is broken. Because of this property, optical SHG has been used recently to study noncentrosymmetric Weyl semimetals (*31*) and odd-parity order in correlated iridates and cuprates (*32, 33*). To completely resolve the structure of $\chi_{ijk}$ for Cd$_2$Re$_2$O$_7$, we used a recently-developed high-speed rotational anisotropy (RA) technique (*34, 35*) that involves focusing a beam of light obliquely onto the surface of a single crystal and measuring variations in the intensity of reflected SHG light as the scattering plane is rotated about the surface normal. By projecting the SHG signal radiated at different scattering plane angles $\phi$ onto a circular locus of points on a stationary 2D detector (Fig. 2A), the experiment can be carried out at very high rotational frequencies (~ 4 Hz), which greatly enhances the sensitivity to small changes in symmetry by averaging over laser fluctuations.

In order to isolate a single-phase domain for detailed RA-SHG study, we carried out three successive stages of microscopy on the natural (111) facet of a Cd$_2$Re$_2$O$_7$ single crystal grown by means of vapor transport [section S1 of (*36*)]. First, white-light microscopy was used to select a smooth area free of surface striations (Fig. 2B). Next, wide-field SHG images [section S2 of (*36*)] were acquired on this area both above and below $T_c$ (Fig. 2C). For $T > T_c$ the SHG signal is dominated by the electric dipole response at the surface ($\chi_{ijk}^S$), where inversion symmetry is necessarily broken. The observed spatial uniformity of the intensity indicates a single cubic structural domain. For $T < T_c$, a strong bulk response ($\chi_{ijk}^B$) develops owing to the inversion-symmetry-breaking tetragonal distortion. Three types of tetragonal domains with sharply-defined linear boundaries are clearly resolved, associated with orientations of the main tetragonal axis along each of the three equivalent cubic lattice directions [section S4 of (*36*)]. Last, we performed scanning RA-SHG measurements [section S3 of (*36*)] within a single tetragonal domain. Because inversion symmetry is spontaneously broken at the phase transition, domains with opposite parity, which we label (+) or (−), will naturally form. Because the sign of $\chi_{ijk}^B$ reverses upon spatial inversion, these parity domains exhibit distinct SHG responses ($\chi_{ijk} =$

$\chi^S_{ijk} \pm \chi^B_{ijk}$) arising from interference between the surface and bulk terms. This in turn produces distinct RA-SHG patterns (Fig. 2D).

To finely resolve the symmetries broken across $T_c$, we performed detailed temperature-dependent RA-SHG measurements on a single domain using different combinations of incoming (in) and outgoing (out) light polarizations, which can be either parallel (P) or perpendicular (S) to the scattering plane [Fig. 2A and section S5 of (*36*)]. Examples of RA-SHG patterns acquired with a S$_{in}$–P$_{out}$ polarization geometry for a selection of temperatures near $T_c$ are shown in Fig. 3. For $T > T_c$, the raw RA-SHG images show disconnected arcs centered at $\phi = 0°$ and every 60° interval (Fig. 3A). A polar plot of SHG intensity versus $\phi$ extracted from the raw data shows that the six intensity peaks are equal in magnitude. This is because the (111) surface of Cd$_2$Re$_2$O$_7$ contains three mirror planes and an axis of three-fold rotational symmetry. By imposing these point group symmetries on $\chi^S_{ijk}$, the number of independent non-zero elements is greatly reduced [section S6 of (*36*)]. The nonlinear polarization from the surface calculated under a S$_{in}$–P$_{out}$ geometry takes the simple form $\chi^S \cos(3\phi)$, where $\chi^S$ is shorthand for the $\chi^S_{yyy}$ tensor element. The resulting RA-SHG intensity pattern, which is proportional to the squared magnitude of the nonlinear polarization, exactly reproduces the high-temperature data and reaffirms its surface origin.

Upon lowering the temperature only slightly (< 1 K) below $T_c$, we observed a dramatic change in the symmetry of the RA-SHG pattern (Fig. 3B) caused by the coherent addition of a bulk SHG contribution to the existing surface signal. Surprisingly, the data cannot be accounted for by the $E_u$ lattice distortion alone. This is already apparent at the qualitative level because the $I\bar{4}m2$ lattice structure preserves mirror symmetry across the vertical $(1\bar{1}0)$ plane (Fig. 1C) whereas the RA-SHG pattern is clearly not symmetric under a $\phi \to -\phi$ transformation. To express this analytically, the SHG susceptibility tensor in the case of $E_u$ order contains only one independent non-zero element: $\chi^{E_u} \equiv \chi_{xyz} = \chi_{xzy} = \chi_{yxz} = \chi_{yzx} = -\chi_{zxy}/2 = -\chi_{zyx}/2$, where the Cartesian coordinates $x$, $y$, and $z$ are chosen to be aligned along the tetragonal $a$, $b$, and $c$ axes, respectively. With a S$_{in}$–P$_{out}$ geometry, this introduces a bulk nonlinear polarization term of the form $\chi^{E_u} \cos(\phi)$ that is even in $\phi$ like the surface term, generating an RA-SHG pattern that is likewise even in $\phi$. This is incompatible with the data and suggests that an additional inversion-symmetry-breaking order parameter emerges together with the $E_u$ structural order parameter below $T_c$.

Among the four other odd-parity irreducible representations of the octahedral point group ($A_{1u}$, $A_{2u}$, $T_{1u}$, and $T_{2u}$), only an order parameter with $T_{2u}$ symmetry can couple to the $E_u$ structural order parameter and produce a RA-SHG pattern that breaks mirror symmetry across the $(1\bar{1}0)$ plane. In the particular case of a multipolar nematic instability, it is the spin texture on the Fermi surface that explicitly breaks this mirror symmetry (Fig. 1A). The SHG susceptibility tensor associated with $T_{2u}$ multipolar nematic order ($Q_{xy} \neq 0$) contains two independent non-zero elements: $\chi^{T_{2u}} \equiv \chi_{xxz} = \chi_{xzx} = -\chi_{yyz} = -\chi_{yzy}$ and $\chi_{zxx} = -\chi_{zyy}$. The absence of any detectable bulk SHG signal in a S$_{in}$–S$_{out}$ geometry [section S5 of (*36*)] imposes the additional constraint $\chi_{zxx} = -2\chi_{xxz}$, reducing the number of independent tensor elements to just one. For a

$S_{in}$–$P_{out}$ geometry, this tensor structure produces a bulk nonlinear polarization of the form $\chi^{T_{2u}}\sin(\phi)$, which taken alone would lead to a RA-SHG pattern that is even in $\phi$. However, a coherent superposition of the $\chi^{T_{2u}}\sin(\phi)$, $\chi^{E_u}\cos(\phi)$, and $\chi^S\cos(3\phi)$ terms, illustrated in the right column of Fig. 3, generates a RA-SHG pattern that breaks $\phi \rightarrow -\phi$ symmetry.

To quantitatively assess the validity of this model, we performed fits to the RA-SHG data in all four polarization geometries simultaneously. We fixed $\chi^S$ at its fitted $T > T_c$ value (Fig. 3A) because it does not measurably change across $T_c$ [section S5 of (36)], leaving the two complex numbers $\chi^{E_u}$ and $\chi^{T_{2u}}$ as the only free parameters. This model provides an excellent and unique fit to the data (Fig. 3), providing further evidence of coupled $T_{2u}$ and $E_u$ order parameters below $T_c$. As the temperature is further cooled to just several kelvin below $T_c$, $\chi^{E_u}$ becomes dominant and a pronounced transformation of the RA-SHG pattern towards a $|\chi^{E_u}\cos(\phi)|^2$ form takes place (Fig. 3C), obscuring the $T_{2u}$ order parameter. The relative faintness of the $T_{2u}$ signal is consistent with a nematic instability that predominantly affects states only near the Fermi level and naturally explains the absence of any detectable $T_{2u}$ distortion by structure-sensitive probes (19–22).

Distinguishing a genuine electronic nematic phase transition from a simple ferrodistortive transition is a well-known experimental challenge (11) because the electronic and structural order parameters are typically coupled and have a concurrent temperature onset, as is the case for $Cd_2Re_2O_7$. The task of disentangling primary from secondary order parameters can be approached by studying the critical exponents $\beta$ of the order parameter temperature scaling law $|1 - T/T_c|^\beta$. SHG is particularly well-suited for this because $\chi_{ijk}$ is linearly proportional to order parameters that are parity-odd [section S7 of (36)]. To obtain the temperature dependence of the $E_u$ and $T_{2u}$ order parameters, we acquired RA-SHG patterns over a series of finely spaced temperatures below $T_c$ and fit them to the model previously described. Because $\cos(3\phi)$, $\cos(\phi)$, and $\sin(\phi)$ are orthogonal functions, the fitted values of $\chi^{E_u}$ and $\chi^{T_{2u}}$ at any given temperature can be determined uniquely. Furthermore, because the $E_u$ and $T_{2u}$ tensors have no elements in common, every bulk SHG response channel $\chi_{ijk}$ couples to only one of the two order parameters. The temperature dependence of $|\chi^{T_{2u}}|$, which is proportional to the $T_{2u}$ order parameter, was extracted from such fits (Fig. 4A). An onset temperature of $T_c \approx 201$ K and a critical exponent of $\beta \approx 1/2$ are obtained from a least-squares fit to the scaling law, which is consistent with the mean-field prediction for a primary order parameter. At temperatures below ~198 K, the $T_{2u}$ response is overwhelmed by the $E_u$ response (Fig. 3C) and can no longer be reliably extracted from the data.

The temperature dependence of $|\chi^{E_u}|$, which is proportional to the $E_u$ order parameter, also exhibits an onset at $T_c \approx 201$ K (Fig. 4B), demonstrating that it is coupled to the $T_{2u}$ order parameter. It has a linear temperature dependence ($\beta \approx 1$) extending over a wide temperature range below $T_c$. This behavior is contrary to that expected of a primary order parameter because critical fluctuations may reduce $\beta$ from its mean-field value but can never increase it. Instead, the $E_u$ structural distortion must be a secondary order parameter. To place this interpretation on firmer theoretical grounds, we used a phenomenological Landau free energy analysis. A system with an odd-parity primary order parameter $\Psi_u$ and a secondary $E_u$ order parameter $\Phi_{E_u}$ is described by the generic Landau free energy expansion

$$F = F_0 - \left(1 - \frac{T}{T_c}\right)\left(a_g \Psi_g^2 + a_u \Psi_u^2\right) + b\Phi_{E_u}^2 - g\Psi_g \Psi_u \Phi_{E_u} + \text{higher order terms,}$$

where $a_g$, $a_u$, $b$, and $g$ are temperature-independent parameters. To realize a linear coupling between $\Psi_u$ and $\Phi_{E_u}$, an additional even-parity primary order parameter $\Psi_g$ that transforms like the product $\Psi_u \Phi_{E_u}$ must be introduced. By construction, minimization of the free energy gives $\Psi_u \propto \Psi_g \propto |1 - T/T_c|^{1/2}$ and $\Phi_{E_u} \propto \Psi_u \Psi_g \propto |1 - T/T_c|$, which exactly reproduces our experimental results. By performing a general symmetry-based analysis of the Landau expansion [section S8 of (*36*)], it is possible to constrain the irreducible representations of $\Psi_u$ and $\Psi_g$ to $T_{2u}$ and $T_{1g}$, respectively. This serves as a strong self-consistency check of our RA-SHG data analysis. The $T_{1g}$ order parameter uncovered by this analysis preserves inversion symmetry and is therefore not detectable with SHG. It is possible that interactions among electrons in Re $t_{2g}$ levels may realize a correlated spin-triplet state with $T_{1g}$ symmetry [section S8 of (*36*)], and nuclear quadrupole resonance measurements have in fact detected a moderate ferromagnetic enhancement below $T_c$ (*37*). Further study, however, is required to firmly establish a microscopic origin of the $T_{1g}$ order.

Our data and analysis reveals the existence of a $T_{2u}$ electronic order in Cd$_2$Re$_2$O$_7$ that drives the 200 K phase transition and induces the $E_u$ lattice distortion as a secondary order parameter. The assignment of the $T_{2u}$ order to a multipolar nematic phase is supported by previous experiments that show only weak $E_u$ structural distortions (*20–22*) accompanied by large electronic anomalies across $T_c$ (*26–29*), the absence of any charge or magnetic order below $T_c$ (*37*), and full agreement with theoretical prediction (*14*). More generally, our results establish a distinct class of odd-parity multipolar electronic nematic phases in spin-orbit coupled correlated metals and demonstrates an experimental strategy for uncovering further realizations of such order. Carefully examining the competing phases in the vicinity of odd-parity nematic order, including the superconducting phase below ~ 1 K in Cd$_2$Re$_2$O$_7$ (*38–40*), may prove fruitful for uncovering other unconventional phases of matter.

**Acknowledgments:**

We thank J. P. Eisenstein, L. Fu, T. Hsieh, P. A. Lee, A. de la Torre, and L. Zhao for useful discussions. RA-SHG experiments were supported by the U.S. Department of Energy under grant DE-SC0010533. Instrumentation for the RA-SHG setup was partially supported by a U.S. Army Research Office Defense University Research Instrumentation Program award under grant W911NF-13-1-0293 and the Alfred P. Sloan Foundation under Grant No. FG-BR2014-027. D.H. also acknowledges funding from the Institute for Quantum Information and Matter, a NSF Physics Frontiers Center (PHY-1125565) with support of the Gordon and Betty Moore Foundation through grant GBMF1250. J.-Q.Y. and D.G.M. were supported by the U.S. Department of Energy, Office of Science, Basic Energy Sciences, Materials Sciences and Engineering Division. Z.Y.Z. acknowledges the Center for Emergent Materials, and NSF Materials Research Science and Engineering Center under grant DMR-1420451. D.H. and D. H. Torchinsky are inventors on U.S. patent application #14/705,831 submitted by the California Institute of Technology, which covers a spectrometer apparatus for the study of the crystallographic and electronic symmetries of crystals and methods of using said apparatus. The data that support the plots within this paper and other findings of this study are available from the corresponding author upon reasonable request.

**Supplementary Materials:**

Materials and Methods

Figs. S1 to S3

References (*41–55*)

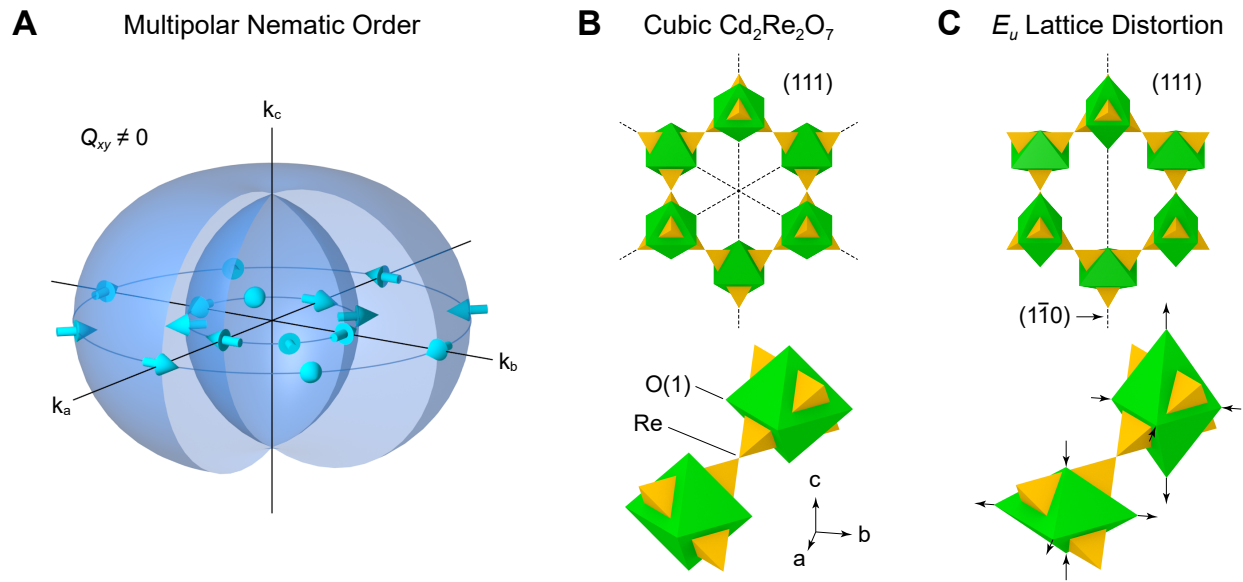

**Fig. 1**. **Illustration of electronic and structural order parameters in Cd$_2$Re$_2$O$_7$.** (**A**) $T_{2u}$ distortion of a spherical Fermi surface induced by a $Q_{xy}$ multipolar nematic order parameter. The Fermi surface is split into two non-degenerate surfaces of opposite spin texture (arrows), with the largest splitting at the equator and zero splitting at the poles. (**B**) The ideal pyrochlore structure of Cd$_2$Re$_2$O$_7$ viewed along the $\langle 111 \rangle$ axis. Only Re (yellow tetrahedra) and O(1) (green octahedra) sublattices are shown. Dashed lines depict mirror planes. For clarity, an enlarged view of two neighboring subunits from an alternative angle is also displayed. (**C**) The effect of the $E_u$ lattice distortion. The vertical dashed line depicts the preserved $(1\bar{1}0)$ mirror plane and arrows show the displacement directions of the O(1) atoms.

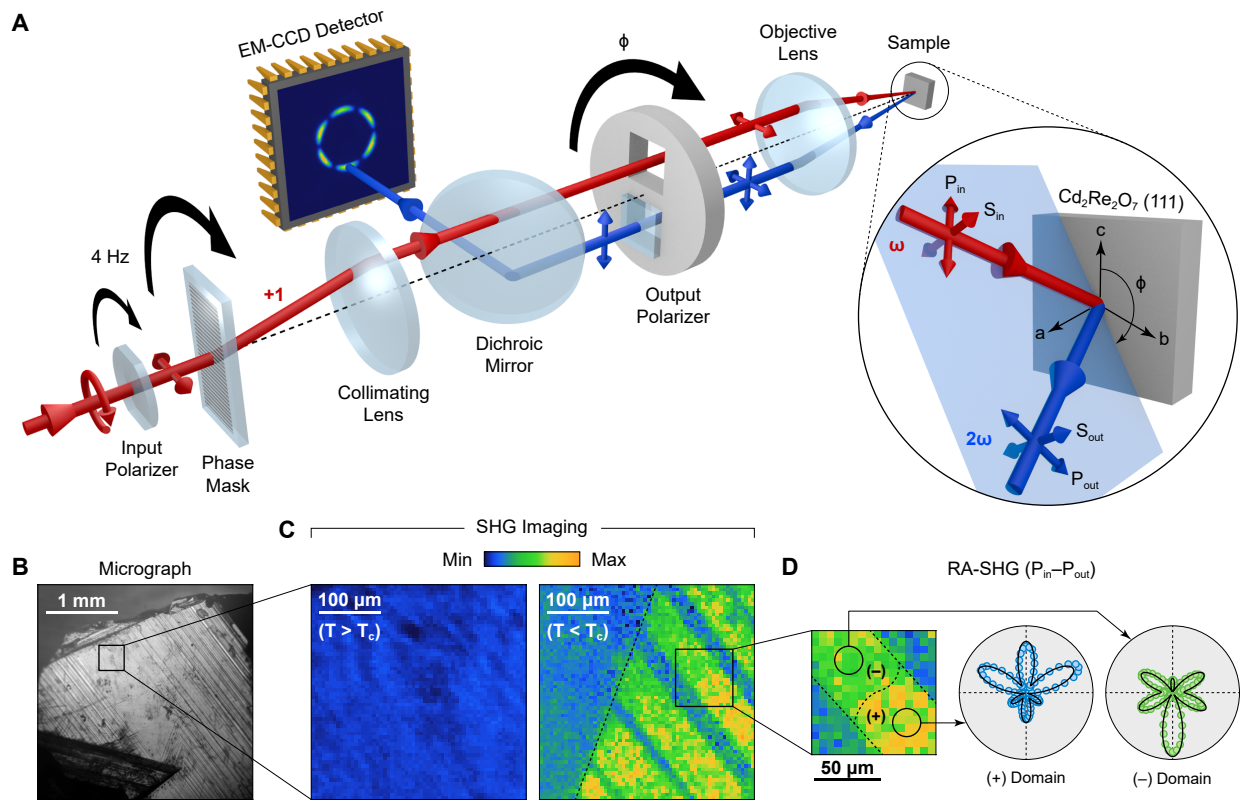

**Fig. 2. Spatially resolved optical SHG anisotropy measurements.** (**A**) Schematic of the RA-SHG setup. A circularly polarized laser beam (red) with center wavelength at 800 nm is sent through a linear polarizer (to select either $P_{in}$ or $S_{in}$ polarization) and onto a phase mask. The +1 diffracted order is isolated and redirected parallel to the optical axis by a collimating lens. After passing through a dichroic mirror, the beam is focused by an objective lens onto the stationary sample to a spot size of ~30 μm. The reflected SHG beam (blue) is recollimated by the objective, passes through a second linear polarizer (to select either $P_{out}$ or $S_{out}$ polarization), and is deflected by the dichroic mirror onto a 2D electron-multiplying charge-coupled device (EM-CCD) camera. The polarizers and phase mask rotate rapidly (black arrows), causing the SHG beam to trace out a circle on the camera as the scattering plane angle $\phi$ changes. (**B**) Optical micrograph of the (111) surface of a $Cd_2Re_2O_7$ single crystal. (**C**) Wide-field SHG image of a striation-free region measured at $T = 210$ K and $T = 150$ K. All three types of tetragonal domains [section S4 of (*36*)] are visible in the low-temperature image. (**D**) Enlarged SHG image of the region over which scanning RA-SHG was performed. $P_{in}$–$P_{out}$ RA-SHG patterns for two opposite parity domains are shown, and an approximate domain boundary is drawn.

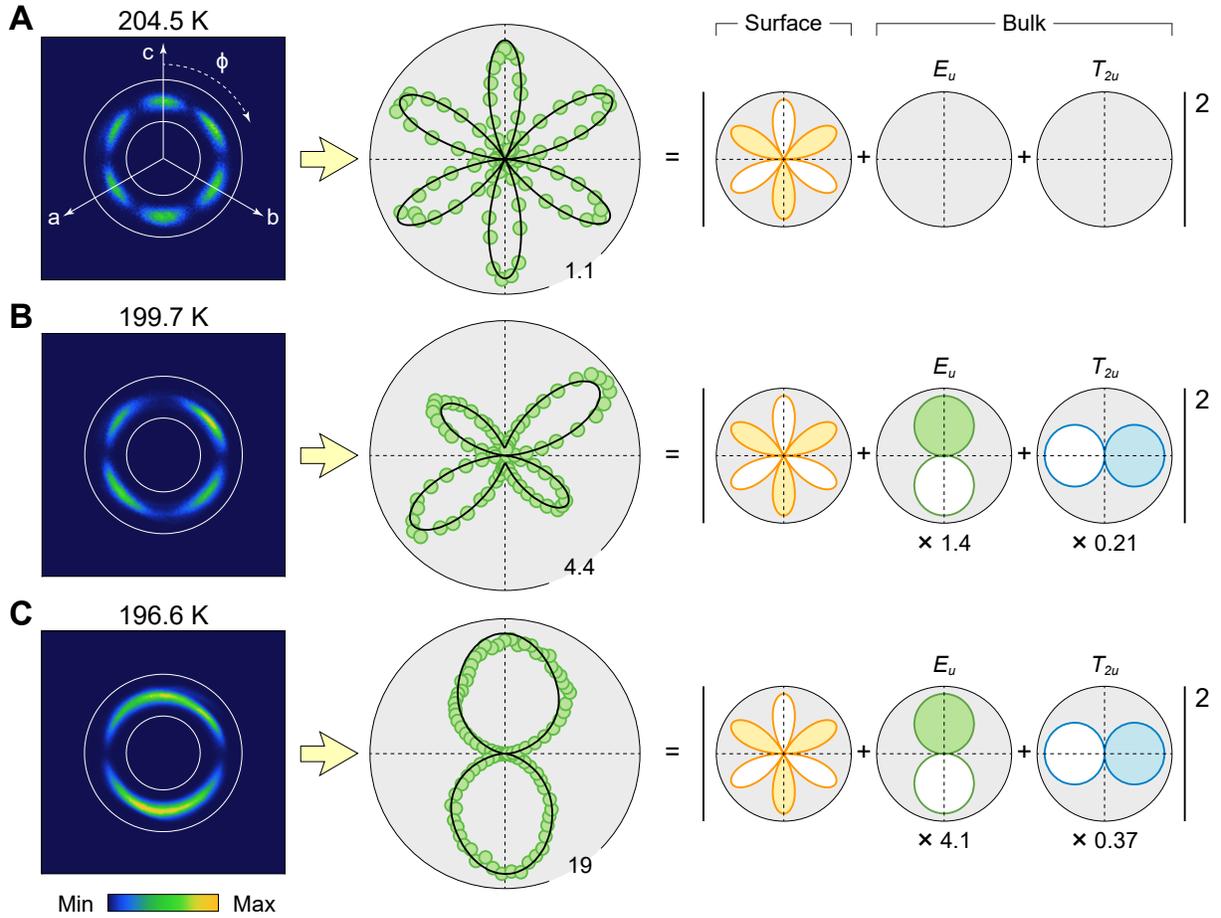

**Fig. 3. Detection of $E_u$ and $T_{2u}$ symmetry breaking by RA-SHG.** Raw RA-SHG images acquired with a $S_{in}$–$P_{out}$ polarization geometry (left column) at temperatures of (**A**) 204.5 K, (**B**) 199.7 K, and (**C**) 196.6 K. Concentric white circles show the radial integration region used to generate the RA patterns (middle column). Numbers on the outer boundaries of the polar plots indicate the intensity scale in units where $\chi^S = 1$. The RA patterns are fit to the squared magnitude of a sum of surface, bulk $E_u$, and bulk $T_{2u}$ polarization terms, as described in the text. Fits are overlaid on the RA patterns (black curves) and each component is illustrated in the right column, where solid petals denote a positive sign and white petals denote a negative sign for the polarization.

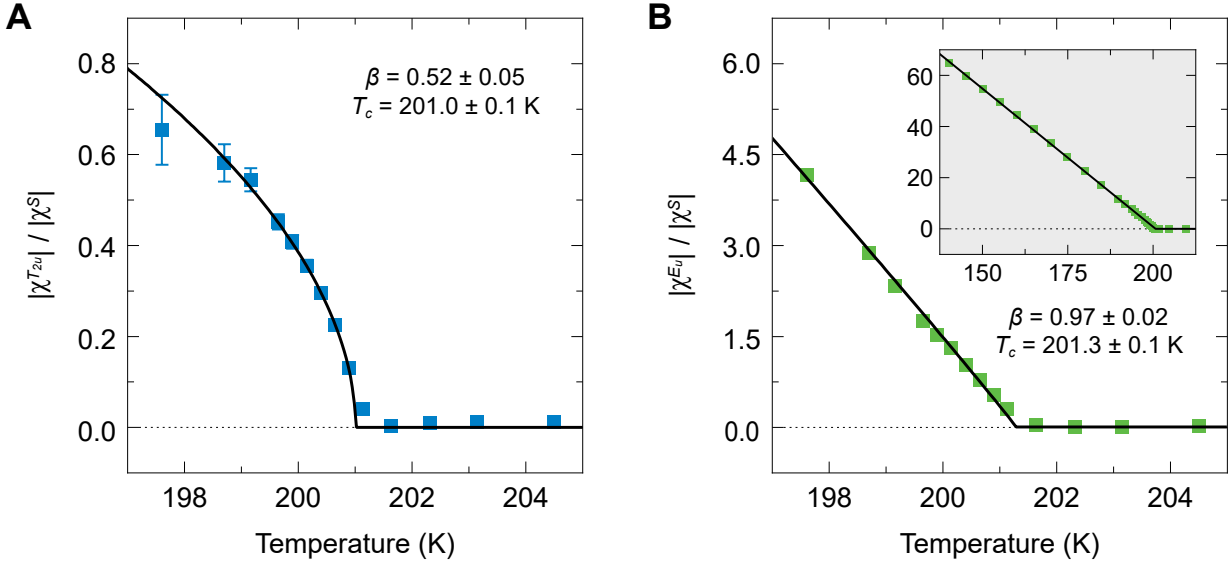

**Fig. 4. Critical exponents of the $T_{2u}$ and $E_u$ order parameters.** Temperature dependence of (**A**) $|\chi^{T_{2u}}|$ and (**B**) $|\chi^{E_u}|$ normalized to the fixed surface contribution $|\chi^S|$. Solid lines show least-squares fits to the scaling law $|1 - T/T_c|^\beta$. Specified uncertainties in the fit parameters are 1 SD. The critical exponent of the $T_{2u}$ order parameter is consistent with the mean-field prediction $\beta = 1/2$, and the critical exponent of the $E_u$ order parameter is consistent with a linear temperature dependence ($\beta = 1$). The inset in (**B**) shows that the linearity of the $E_u$ secondary structural order persists over a wide temperature range below $T_c$.

# Supplementary Materials for

# A parity-breaking electronic nematic phase transition in the spin-orbit coupled metal $Cd_2Re_2O_7$

J. W. Harter, Z. Y. Zhao, J.-Q. Yan, D. G. Mandrus, D. Hsieh*

*Corresponding author. Email: dhsieh@caltech.edu

**This PDF file includes:**

Materials and Methods
Figs. S1 to S3

**Materials and Methods**

**S1. Sample Growth and Characterization**

*Sample Growth*

Single crystals of $Cd_2Re_2O_7$ were grown by vapor transport (*41*). X-ray diffraction measurements were performed on pulverized single crystals using a PANalytical X'Pert Pro powder x-ray diffractometer with Cu Kα radiation. No impurity peaks were observed. An elemental analysis was performed using a Hitachi TM-3000 scanning electron microscope equipped with a Bruker QUANTAX 70 energy dispersive x-ray system. The analysis confirmed an equal amount of Cd and Re within the resolution of the instrument. Magnetic susceptibility measurements were performed using a Quantum Design Magnetic Property Measurement System at temperatures ranging from 2 to 350 K. The results indicated high quality crystals without the presence of $ReO_2$ inclusions.

*Stoichiometric Origin of SHG Heterogeneity*

Wide-field SHG microscopy uncovered an inhomogeneous bulk response of the sample consisting of bright and dark regions. The boundaries between these regions are curved, tend to be near the edges of the sample, appear to be unaffected by structural domains, and have the same size scale as the single crystal itself. These facts strongly suggest chemical inhomogeneity during sample growth as a likely cause. To verify this hypothesis, we performed energy-dispersive x-ray spectroscopy (EDS) using an Oxford X-MaxN silicon drift detector attached to a Zeiss LEO 1550VP scanning electron microscope (SEM). This technique allows us to measure the spatially-resolved chemical composition of the sample near its surface. We collected EDS spectra at eight points—four inside a dark region and four inside a bright region—near a well-defined feature on the (111) sample surface, as shown in Fig. S1.

A typical EDS spectrum is shown in Fig. S1C. Peaks from O, Cd, and Re atoms can be identified, and fitting the spectrum allows us to infer the atomic percentages of the atoms at each sample location. Our results are summarized in the following table:

| Location | Region | O atomic % | Cd atomic % | Re atomic % | Cd/Re ratio |
|---|---|---|---|---|---|
| 1 | Dark | 61.85 | 19.17 | 18.98 | 1.0100 |
| 2 | Bright | 64.31 | 18.14 | 17.55 | 1.0336 |
| 3 | Bright | 62.39 | 19.08 | 18.53 | 1.0297 |
| 4 | Dark | 62.75 | 18.82 | 18.43 | 1.0212 |
| 5 | Bright | 62.52 | 19.08 | 18.40 | 1.0370 |
| 6 | Dark | 62.09 | 19.06 | 18.85 | 1.0111 |
| 7 | Bright | 60.86 | 19.73 | 19.41 | 1.0165 |
| 8 | Dark | 61.16 | 19.56 | 19.28 | 1.0145 |

Variability of oxygen stoichiometry is a known issue in the growth of $Cd_2Re_2O_7$ (*41*) and is a prime candidate for the origin of the heterogeneity, but the oxygen atomic percentage does not appear to correlate with the bright and dark regions within the resolution of our experiment. We do detect, however, a difference in the average Cd/Re atomic ratio: $1.029 \pm 0.009$ in the bright region and $1.014 \pm 0.005$ in the dark region. A Student's *t*-test of the data shows that the difference is statistically significant ($p = 0.027$). The measured deviations of the Cd/Re ratio from 1 are within the *absolute* accuracy of the EDS instrument, so we cannot determine which of the two regions is more stoichiometric.

In oxide pyrochlores with chemical formula $A_2B_2O_7$, the A and B cations can intermix under certain conditions and may even form a stable disordered phase (*42*). Indeed, in so-called "stuffed spin-ice" pyrochlores, the A/B atomic ratio can exceed 1.9 (*43*). Given our various observations as a whole, the most likely scenario is that inhomogeneous growth conditions cause Re atoms, with a much smaller ionic radius, to occupy some of the Cd sites in the darker regions of the sample. This will cause the nanoscale crystallinity to be slightly worse there (and may result in some localized metallic doping). It is then natural to expect the order parameter to be weaker in the more disordered regions of the sample, as we observe. For this reason, our study focuses on bright regions, where the bulk order parameters are larger with respect to the surface term.

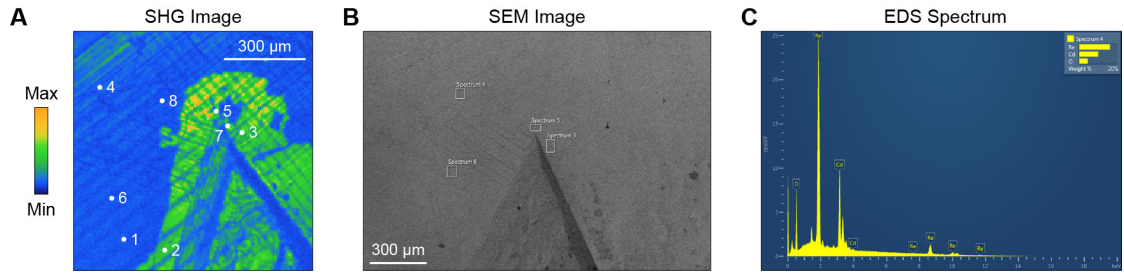

**Fig. S1. Spatial mapping of sample stoichiometry.** (**A**) SHG image of the sample at 150 K near a well-defined feature on the surface. Eight sample points in bright and dark regions measured with EDS are indicated. (**B**) SEM image showing four of the rectangular regions used for EDS measurements. (**C**) Typical EDS spectrum, with x-ray peaks identified for O, Cd, and Re atoms.

## S2. Wide-Field SHG Microscopy

Wide-field SHG imaging was performed using ultrashort 100 fs optical pulses with a center wavelength of 800 nm produced at a 100 kHz repetition rate by a regeneratively-amplified Ti:sapphire laser system (Coherent Vitara-S and RegA 9050). The beam illuminated the entire (111) sample facet at an oblique 20° angle of incidence and with a fluence below 30 μJ cm$^{-2}$. The reflected second-harmonic light was selected with a 400 nm narrow bandpass filter and linear polarizer, collected by an objective lens, and focused onto an EM-CCD camera (Andor iXon Ultra 897). Each pixel in the image captured a $6 \times 6$ μm$^2$ area on the sample surface. Upon temperature cycling, we did not observe any movement of tetragonal domains, suggesting that they are pinned by static strain fields and other crystallographic defects.

## S3. RA-SHG Measurements

RA-SHG measurements were performed with the same 800 nm light source as that used for the microscopy measurements. Both the sample and detector remained fixed. Rotation of the scattering plane was achieved by mechanically spinning transmissive optical elements about the central beam axis. A detailed description of the novel RA-SHG apparatus used can be found in Ref. *34*. The laser was obliquely incident on the sample with a fixed 10° angle between the beam and the sample surface normal. The fluence of the beam was maintained at 600 µJ cm$^{-2}$, with no noticeable degradation of the sample with time. The laser spot size on the sample was 30 µm full-width at half-maximum. Reflected second-harmonic light at 400 nm was selected with a narrow bandpass filter and measured with a two-dimensional EM-CCD (Andor iXon Ultra 897). Each complete RA pattern was acquired with a 30 s exposure time. Samples were measured in an optical cryostat with a vacuum pressure below 10$^{-6}$ Torr. We point out that an 800 nm wavelength ($\hbar\omega = 1.5$ eV) is resonant with interband transitions between occupied O 2$p$ and unoccupied Re 5$d$ states in Cd$_2$Re$_2$O$_7$ (*44–46*), significantly enhancing the SHG response of the crystal at this wavelength. We did not observe substantial steady-state heating of the sample based on the close agreement between $T_c$ measured using SHG and $T_c$ previously reported in the literature (*26*). This is consistent with a calculation of the heating amount (~ 2 K) using a thermal conductivity of 3 W m$^{-1}$ K$^{-1}$ (*47*).

## S4. Identification of All Six Domain Types

A cubic-to-tetragonal distortion will in general result in three types of domains, with each aligned along one of the three equivalent cubic lattice directions in the crystal. In addition, a parity-breaking distortion can occur in two separate ways related by inversion symmetry. Aided by SHG imaging, we are indeed able to find all six possible types of structural domains (three tetragonal × two parity) on the (111) sample surface. Performing RA-SHG measurements at each of the six domains yields distinct RA patterns, allowing us to fully distinguish them, as shown in Fig. S2.

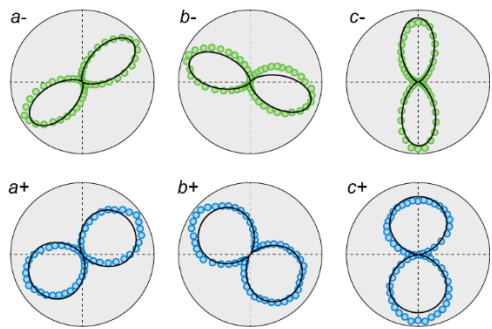

**Fig. S2. Identification of domains via RA-SHG.** Normalized S$_{in}$–P$_{out}$ RA-SHG patterns measured at 150 K at various locations on the sample surface showing all six possible structural domains. The three types of tetragonal domains (*a*, *b*, *c*) are distinguished by the directions of the RA lobes, which differ by 120°. The two types of parity domains (+, –) are distinguished by the shapes of the RA lobes, which differ because of interference between bulk and surface SHG.

## S5. RA-SHG Data for All Four Polarization Geometries

To minimize systematic errors, we use all four linear polarization geometries when fitting the nonlinear susceptibility to the data. At each temperature, all four RA patterns ($P_{in}$–$P_{out}$, $P_{in}$–$S_{out}$, $S_{in}$–$P_{out}$, and $S_{in}$–$S_{out}$) are measured and fit simultaneously with the *same* set of free parameters. Fig. S3 displays RA patterns and resulting fits for a representative selection of temperatures taken at (+) and (–) domains in dark regions of the sample, as well as a (–) domain in a bright region of the sample. Notably, there is no detectible bulk response for $S_{in}$–$S_{out}$. We use this fact to constrain some of the bulk susceptibility components, as discussed in S6 below. Parity domains can be clearly distinguished from each other by the anisotropy resulting from interference with the surface SHG component. Because the bulk response is much stronger relative to the surface in the bright regions, our sensitivity to bulk symmetry breaking near $T_c$ is increased there.

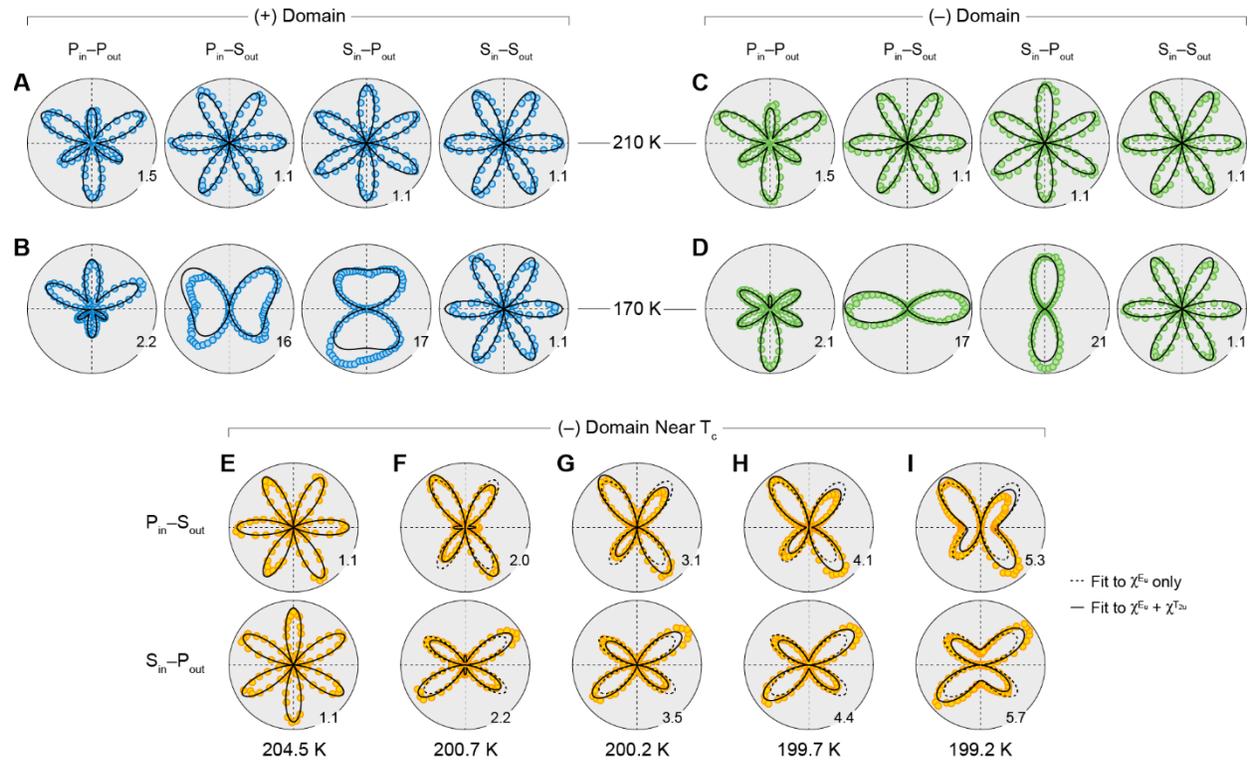

**Fig. S3. RA-SHG domain contrast. (A)-(D)** RA-SHG data with all four polarization geometries for a (+) domain (A, B) and a (–) domain (C, D) taken above $T_c$ at 210 K (A and C) and below $T_c$ at 170 K (B and D). At high temperatures, only the surface SHG contributes and the two domains are identical. Below $T_c$, interference between the static surface and growing bulk SHG contributions generates RA contrast between the parity domains. **(E)-(I)** $P_{in}$–$S_{out}$ and $S_{in}$–$P_{out}$ RA-SHG data taken at 204.5 K (E), 200.7 K (F), 200.2 K (G), 199.7 K (H), and 199.2 K (I). The loss of horizontal reflection symmetry in the RA curves is clearly detected below $T_c$ but is not captured by fitting the data to a $\bar{4}2m$ bulk point group (dashed curves). A $\bar{4}$ bulk point group, however, fits the data well (solid curves).

## S6. Surface and Bulk Susceptibility Tensors

In our RA-SHG measurements, we observe a static, temperature-independent SHG component from inversion symmetry breaking at the sample surface. To isolate this component from the bulk response, we split the second-order optical susceptibility tensor that we measure into surface and bulk parts. Optical susceptibilities are complex. Our experiments, however, are only sensitive to relative phase differences between the susceptibility components, so we define the surface part to be strictly real but allow a complex phase for the bulk: $\chi_{ijk}^{measured} = \chi_{ijk}^{S} + \chi_{ijk}^{B}$.

In the cubic phase, the (111) surface of $Cd_2Re_2O_7$ contains three mirror planes and a three-fold rotation axis and is represented by point group $3m$. This symmetry group, together with SHG permutation symmetry, restricts the possible nonzero tensor elements of $\chi_{ijk}^{S}$ to $\chi_{xxz} = \chi_{xzx} = \chi_{yyz} = \chi_{yzy}$, $\chi_{zxx} = \chi_{zyy}$, $\chi_{yxx} = \chi_{xxy} = \chi_{xyx} = -\chi_{yyy}$, and $\chi_{zzz}$, where coordinate axes are chosen so that one of the mirror planes is perpendicular to the $x$-axis and the surface normal is parallel to the $z$-axis (30). At high temperatures, where only $\chi_{ijk}^{S}$ is nonzero, we find best agreement with our data when $\chi_{xxz} = -0.9$, $\chi_{zxx} = 0$, and $\chi_{zzz} = 200$, in units where $\chi_{yyy} = 1$. We fix these values during our low-temperature fits and do not let them vary. In the $y$-axis labels of Fig. 4, "$|\chi^S|$" is shorthand for the value of the $\chi_{yyy}$ surface tensor element.

Our experiments show that $\chi_{ijk}^{B}$ cannot be faithfully represented by a point group of higher symmetry than $\bar{4}$, whose nonzero tensor elements are $\chi_{xxz} = \chi_{xzx} = -\chi_{yyz} = -\chi_{yzy}$, $\chi_{xyz} = \chi_{xzy} = \chi_{yxz} = \chi_{yzx}$, $\chi_{zxx} = -\chi_{zyy}$, and $\chi_{zxy} = \chi_{zyx}$ (30). Here, the coordinate axes $x$, $y$, and $z$ are chosen to be aligned with the cubic crystallographic directions $a$, $b$, and $c$, respectively. As discussed in the main text, it is physically more appropriate to split this general bulk tensor into two parts, labeled $\chi_{ijk}^{E_u}$ and $\chi_{ijk}^{T_{2u}}$, that transform like point groups $\bar{4}2m$ (the isotropy group induced by $E_u$ order) and $\bar{4}m2$ (the isotropy group induced by $T_{2u}$ order), respectively. Strictly speaking, these point groups are synonymous, differing by a trivial 45° rotation about the $\bar{4}$ axis. Within a *space* group, however, the cubic crystallographic axes define fixed directions in space and the two become inequivalent. This decomposition of the $\bar{4}$ tensor can be performed without loss of generality and unambiguously because $\bar{4}2m$ tensors contain only $\chi_{xyz} = \chi_{xzy} = \chi_{yxz} = \chi_{yzx}$ and $\chi_{zxy} = \chi_{zyx}$ and $\bar{4}m2$ tensors contain only $\chi_{xxz} = \chi_{xzx} = -\chi_{yyz} = -\chi_{yzy}$ and $\chi_{zxx} = -\chi_{zyy}$. In particular, no tensor elements are shared between the two point groups.

We observe an additional symmetry in the $\chi_{ijk}^{B}$ tensor: $\chi_{zxy} = -2\chi_{xyz}$ and $\chi_{zxx} = -2\chi_{xxz}$. Experimentally, this symmetry is manifest in the complete absence of bulk SHG for the $S_{in}$–$S_{out}$ polarization geometry (see S5 above). For $\chi_{ijk}^{E_u}$, this constraint is directly related to the fact that a general $\bar{4}2m$ tensor transforms like the direct sum $E_u \oplus A_{2u}$. By setting $\chi_{zxy} = -2\chi_{xyz}$, the $A_{2u}$ part is projected out of the susceptibility tensor. For $\chi_{ijk}^{T_{2u}}$, the constraint $\chi_{zxx} = -2\chi_{xxz}$ is not imposed by $T_{2u}$ macroscopic symmetry considerations alone, but may follow from symmetries in the microscopic Hamiltonian. A further discussion of these additional symmetries can be found in S7 below.

In the end, only two independent complex parameters are allowed to vary with temperature in the susceptibility model that we employ: $\chi_{xxz}$ and $\chi_{xyz}$. In the y-axis labels of Fig. 4, "$|\chi^{E_u}|$" is shorthand for the magnitude of the $\chi_{xyz}$ bulk tensor element and "$|\chi^{T_{2u}}|$" is shorthand for the magnitude of the $\chi_{xxz}$ bulk tensor element.

## S7. Linear Proportionality of Bulk Susceptibility and Odd-Parity Order Parameters

For a parity-breaking phase transition, we expect the bulk second-order optical susceptibility tensor to be linearly proportional to the order parameter for temperatures sufficiently close to $T_c$. The susceptibility, like the order parameter, is zero in the high-temperature centrosymmetric phase and nonzero in the low-temperature noncentrosymmetric phase. Furthermore, if the order parameter reverses sign, for example in different parity domains related by spatial inversion, so does the susceptibility. We have observed this sign reversal in our measurements and it allows us to distinguish (+) and (−) domains, as shown in S5 above. This means that a generic series expansion of the susceptibility in the order parameter will include only odd powers, with a linear term being the first allowed.

A more rigorous proof that the nonlinear optical susceptibility is linearly proportional to the odd-parity order parameters may be obtained by expanding the Landau free energy up to lowest order in powers of components of the electric polarization **P** (a polar vector transforming like $T_{1u}$) and the odd-parity order parameters $\Psi_u$ (representing $T_{2u}$ order) and $\Phi$ (representing $E_u$ order) (*48*):

$$F(\mathbf{P}) = F_0 + \frac{a}{2} P_i P_i - E_i P_i - b \Psi_u F_1(\mathbf{P}) - c\Phi F_2(\mathbf{P}) + \cdots,$$
$$F_1(\mathbf{P}) = (P_x^2 - P_y^2) P_z [1 - dP_z^2/(P_x^2 + P_y^2 + P_z^2)],$$
$$F_2(\mathbf{P}) = P_x P_y P_z [1 - 3P_z^2/(P_x^2 + P_y^2 + P_z^2)],$$

where summation over repeated indices is implied, **E** is the externally applied optical electric field, $a > 0$ because the system is not ferroelectric, $b$ and $c$ are constants coupling the electric polarization to the $T_{2u}$ and $E_u$ order parameters, respectively, and $d$ is a dimensionless constant determined by the microscopic Hamiltonian of the system. Minimizing the free energy with respect to variation of the components of the electric polarization gives

$$P_i = \frac{1}{a} E_i + \left(\frac{b\Psi_u}{a}\right) \frac{\partial F_1}{\partial P_i} + \left(\frac{c\Phi}{a}\right) \frac{\partial F_2}{\partial P_i}.$$

By noting that $P_i = \chi_{ij} E_j + \chi_{ijk} E_j E_k + \cdots$, we may write

$$\chi_{ijk} = \frac{1}{2} \frac{\partial^2 P_i}{\partial E_j \partial E_k}\bigg|_{E=0} = \frac{1}{2a^2} \lim_{P_i \to 0} \lim_{P_j \to 0} \lim_{P_k \to 0} \frac{\partial^2 P_i}{\partial P_j \partial P_k}.$$

This formula gives $\chi_{xyz} = \chi_{xzy} = \chi_{yxz} = \chi_{yzx} = -\chi_{zxy}/2 = -\chi_{zyx}/2 = c\Phi/2a^3$, explicitly showing the linear proportionality of $\chi^{E_u}$ and $\Phi$ and confirming the additional symmetry constraint $\chi_{zxy} = -2\chi_{xyz}$ discussed in S6 above. We also obtain $\chi_{xxz} = \chi_{xzx} = -\chi_{yyz} = -\chi_{yzy} = b\Psi_u/a^3$ and $\chi_{zxx} = -\chi_{zyy} = (1-d)b\Psi_u/a^3$, showing the proportionality of $\chi^{T_{2u}}$ and $\Psi_u$.

Interestingly, the empirical relation $\chi_{zxx} = -2\chi_{xxz}$ that we observe in our measurements is achieved if $d = 3$, which is likely the result of a microscopic symmetry in the Hamiltonian of the system and offers a strong constraint on possible theories of the multipolar nematic phase in $Cd_2Re_2O_7$.

## S8. Landau Theory Symmetry Analysis

Within Landau's theory of second-order phase transitions, an order parameter $\Phi$ that grows linearly with temperature in the low-symmetry phase, $\Phi \propto (1 - T/T_c)$ for $T < T_c$, must be secondary, induced by a coupling to a primary order parameter $\Psi$ with mean field temperature dependence $\Psi \propto \sqrt{1 - T/T_c}$ (*49*). The so-called "faintness index" $n \geq 2$ encodes the exponent of the primary order parameter that couples to the secondary order parameter in the Landau free energy, producing an invariant term of the form $\Psi^n \Phi$ (*50*). Ferroic transitions that occur under these conditions are labeled "improper" (*49–51*). In particular, for a secondary order parameter with linear temperature dependence, $n = 2$.

A parity-breaking secondary order parameter with $n = 2$ implies the existence of exactly two coupled primary order parameters, one that breaks inversion symmetry and one that preserves it. This is because a term like $\Psi^2 \Phi$ is not invariant under the parity operation if $\Phi$ breaks inversion symmetry. In analogy with Mulliken notation, let us call the parity-even order parameter $\Psi_g$ and the parity-odd order parameter $\Psi_u$. Then a coupling term like $\Psi_g \Psi_u \Phi$ is invariant with respect to inversion symmetry and is allowed in the Landau free energy

$$F(T) = F_0 - \frac{a}{2}\left(1 - \frac{T}{T_c}\right)\left(\Psi_g^2 + \Psi_u^2\right) + \frac{b}{2}\Phi^2 - g\Psi_g\Psi_u\Phi + \frac{c}{4}\left(\Psi_g^4 + \Psi_u^4\right) + \cdots,$$

where $a$, $b$, and $c$ are positive constants to guarantee a stable solution and $g$ controls the strength of the coupling between the primary and secondary order parameters. For convenience, we have taken the expansion coefficients of $\Psi_g$ and $\Psi_u$ to be identical ($a_g = a_u$) and have ignored higher-order terms in the expansion; such simplifications are irrelevant to the arguments that follow. It should be noted, however, that the identical temperature dependence of the leading order coefficients for the primary order parameters—positive above $T_c$ and negative below $T_c$—ensures that $\Psi_g$ and $\Psi_u$ become nonzero at the same temperature. Although not strictly guaranteed by theoretical arguments alone, such a condition is realized in practice if the coupled primary orders have the same physical origin. For example, the improper ferroelectric transitions observed in $PbTiO_3/SrTiO_3$ superlattices and $(Ca,Sr)_3Ti_2O_7$ are driven by the simultaneous freezing of two phonon modes of different one-dimensional irreducible representations, one preserving inversion symmetry and the other breaking it (*52, 53*).

Minimization of the free energy with respect to variation of the order parameters gives

$$\Psi_g(T) = \Psi_u(T) = \sqrt{\frac{ab}{bc-g^2}}\sqrt{1 - \frac{T}{T_c}}, \quad \Phi(T) = \left(\frac{ag}{bc-g^2}\right)\left(1 - \frac{T}{T_c}\right).$$

By construction, we obtain a linear temperature dependence for the secondary order parameter $\Phi$. Higher-order terms in the free energy expansion or differing coefficients for $\Psi_g$ and $\Psi_u$ will, in general, change the constant factors in the above expressions but will not change the temperature dependence near $T_c$. This analysis shows that in $Cd_2Re_2O_7$, the $E_u$ structural distortion occurring at the 200 K phase transition (here represented by $\Phi$) must be driven by a pair of primary order parameters $\Psi_g$ and $\Psi_u$.

To proceed further, we impose four rules constraining the irreducible representations of the primary order parameters $\Psi_g$ and $\Psi_u$.

- **Rule #1:** The Landau condition must hold, which states that for a second-order phase transition, no third-degree term can appear in the free energy expansion. This eliminates the $A_{1g}$, $E_g$, and $T_{2g}$ irreducible representations for $\Psi_g$ *(54)*.
- **Rule #2:** $\Psi_u$ cannot transform like $E_u$. Otherwise, the term $\Psi_u\Phi$ would occur in the free energy and the transition would not be improper ($\Phi$ would have a square-root temperature dependence).
- **Rule #3:** In order for $\Psi_g\Psi_u\Phi$ to be an allowed invariant in the free energy, the symmetric direct product of the irreducible representations for $\Psi_g$ and $\Psi_u$ must contain $E_u$.

These first three rules immediately imply that $\Psi_g$ must transform like $T_{1g}$, and leave only two possibilities for the irreducible representation of $\Psi_u$: $T_{1u}$ or $T_{2u}$. We now invoke the so-called "maximal isotropy group condition" *(51)*, which states that because $\Psi_g$ and $\Psi_u$ transform like multidimensional irreducible representations, their directions in order parameter space must correspond to a maximal isotropy subgroup. Thus, we restrict our attention to the tetragonal isotropy subgroups that result from order parameter directions aligned with one of the cubic crystallographic axes. The $E_u$ structural order parameter in $Cd_2Re_2O_7$ is two-dimensional, but only the $E_u^{(2)}$ partner is observed by structure-sensitive probes. This fact leads to a final rule that allows us to resolve the ambiguity in the symmetry of the $\Psi_u$ order parameter.

- **Rule #4:** In the free energy expansion, $\Psi_g\Psi_u$ cannot couple to the structural partner $E_u^{(1)}$. Otherwise, it would be finite below $T_c$ and detected by experiment.

Only one pair of order parameter symmetries is consistent with this final rule: $\Psi_g$ transforms according to $T_{1g}$ in the $\langle 001 \rangle$ direction [isotropy subgroup $I4_1/a$ *(48)*] and $\Psi_u$ transforms according to $T_{2u}$ in the $\langle 001 \rangle$ direction [isotropy subgroup $I\bar{4}2d$ *(48)*]. The maximal common subgroup of the order parameter isotropy groups is $I\bar{4}$, which is the true space group of $Cd_2Re_2O_7$ in the low temperature phase.

We emphasize that this analysis, which is based solely on the observed linear temperature dependence of the $E_u$ order parameter, necessitates the existence of a $T_{2u}$ primary order parameter and therefore serves as a strong self-consistency check of our observations. This symmetry-based evidence of $T_{2u}$ order also predicts the existence of a coupled $T_{1g}$ primary order. The apparent coupling between the $T_{2u}$ and $T_{1g}$ primary orders suggests that they have a similar physical origin, and we now discuss a possible mechanism for $T_{1g}$ electronic order in Cd$_2$Re$_2$O$_7$ ($T_{1g}$ transforms like a rotation, such as an angular momentum). Re$^{5+}$ ions have a 5$d^2$ valence configuration. Density functional theory calculations show that the octahedral crystal field strongly splits the 5$d$ orbitals, pushing the unoccupied $e_g$ states ~ 5 eV higher in energy than the occupied $t_{2g}$ states (*44–46*). If we consider a single Re$^{5+}$ ion in isolation, the spatial part of the two-valence-electron wave function will transform according to the direct product $t_{2g} \otimes t_{2g} = A_{1g} + E_g + [T_{1g}] + T_{2g}$, where the brackets denote the antisymmetric combination. According to Hund's first rule, electron interactions will favor the spin triplet state with term symbol $^3T_{1g}$ (*55*). Although the preceding argument applies only to isolated ions, we anticipate that within Cd$_2$Re$_2$O$_7$ interactions among electrons in the Re $t_{2g}$ levels may favor a correlated triplet state with $T_{1g}$ symmetry. A nuclear quadrupole resonance experiment indeed detected a moderate ferromagnetic enhancement, but no static magnetic order was observed (*37*), calling a simple spin triplet state into question. More theoretical work is necessary to identify a suitable physical mechanism for $T_{1g}$ electronic order.